\documentclass[10pt,twocolumn,dvips]{revtex4}
\usepackage[english]{babel}
\usepackage{epsfig}
\usepackage{graphics} 
\usepackage{amsmath}
\begin{document}
\title{Correlation Length Exponent in the Three-Dimensional Fuse Network}
\author{Thomas Ramstad, Jan {\O}.\ H.\ Bakke, Johannes Bjelland, 
Torunn Stranden and Alex Hansen}
\affiliation{Department of Physics, Norwegian University of Science and 
Technology, N--7491 Trondheim, Norway}
\begin{abstract}
We present numerical measurements of the critical correlation length 
exponent $\nu$ in the three-dimensional fuse model.  Using sufficiently
broad threshold distributions to ensure that the system is the 
strong-disorder regime, we determine $\nu$ to be  $\nu = 0.86 \pm 0.06$
based on analyzing the fluctuations of the survival probability.  The value
we find for $\nu$ is very close to the percolation value $0.88$ and we
propose that the three-dimensional fuse model is in the universality class
of ordinary percolation.
\end{abstract}
\maketitle

It is already twenty years since the publication of the first experimental 
evidence of scaling in the morphology of brittle fractures \cite{mpp84}. 
About seven years later it was proposed that not only is there scaling, but
the scaling properties are {\it universal,\/} in the sense that they do not
depend on material properties \cite{blp90,b97}.  There is now mounting
evidence for this hypothesis, which may be expressed as the scaling invariance
\begin{equation}
\label{eq1}
\pi(z;x,y)  = \lambda^{\zeta} \pi(\lambda^{\zeta} z; \lambda x, \lambda y)\;,
\end{equation}
of $\pi$, which is the probability density that at position $(x,y)$ in the 
average fracture plane, the fracture is at height $z$ given that it is at 
$z=0$ at $(0,0)$, with $\zeta$ as the universal roughness exponent having a 
value very close to 0.80 for a large class of materials.  One experimentally
important consequence of this scaling is that the average fracture width $w$ 
scales as
\begin{equation}
\label{wlxz}
w\sim L^\zeta\;,
\end{equation}
where $L$ is the linear size of the average fracture plane.

Ever since the proposal of universality, it has remained a theoretical 
challenge to explain this value.  Recently, it was suggested by Hansen
and Schmittbuhl that it has its origin in the fracture process being a
a correlated percolation process \cite{hs03}.  The essence of the argument
is based on existence of a localization length $l$ and a correlation length
$\xi$ that grows during the breakdown process.  The localization length
depends on the disorder in the material: Stronger disorder means larger
localization length.  Whether the localization length diverges for large but
finite disorder or it only reaches this limit for infinite disorder is at
present not known.  However, mean field arguments suggest that the former
scenario is the correct one \cite{hhr91}.  For correlation lengths $\xi$
much smaller than the localization length $l$, Hansen and Schmittbuhl 
\cite{hs03} assumed a relation 
\begin{equation}
\xi\sim |p-p_c|^{-\nu}\;,
\label{xip}
\end{equation}
where $p$ is the local damage density and $p_c$ is the damage density at
failure.  This relation is taken directly from percolation theory.  The
reason it is only valid for large localization lengths $l$ is that $p$ is 
assumed to be spatially stationary (meaning that the statistical distribution 
of $p$-values is independent of position).  The
correlation length exponent $\nu$ has the value 4/3 in two-dimensional
percolation and 0.88 in three-dimensional percolation \cite{sa92}.  
It is by no means given that $\nu$ should be the same in the brittle 
fracture problem --- and Toussaint and Pride suggest that it is equal to
2 \cite{tp02}.  However, it was suggested by Hansen and Schmittbuhl that the 
two-dimensional fuse model has $\nu=4/3$ placing it in the same universality 
class as two-dimensional percolation.  When the correlation length approaches
the localization length $l$, gradients develop in the damage --- $p$ can no
longer be regarded as spatially stationary --- and using 
arguments from gradient percolation \cite{srg85}, Hansen and Schmittbuhl
suggested the relation
\begin{equation}
\label{zetanu}
\zeta=\frac{2\nu}{1+2\nu}\;.
\end{equation}
With $\nu=4/3$ for the two-dimensional fuse model, this leads to 
$\zeta=8/11\approx 0.73$.  Recent numerical calculations gives 
$\zeta=0.74\pm0.03$ \cite{bbrsh03}.  

\begin{figure}[t!]
\centering
\epsfig{file = 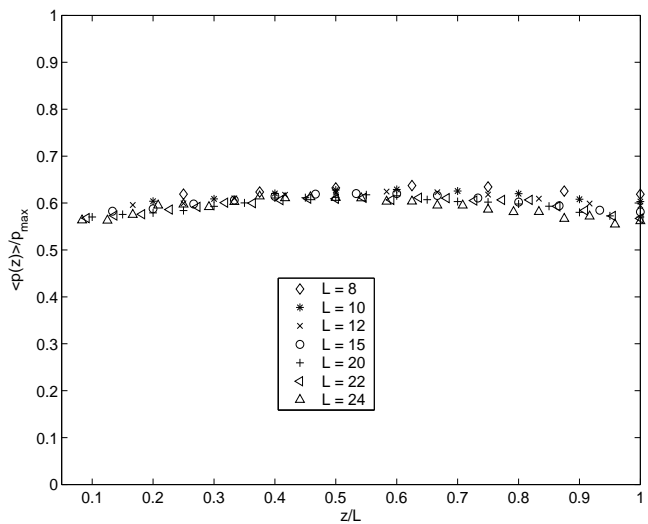, height = 7.0 cm, width = \linewidth, clip=}
\caption{Normalized histogram where each bin is averaged over a plane 
orthogonal to the (1,1,1)-direction for $D=10$.}
\label{fig1}
\end{figure}

Recently, Kumar et al.\ \cite{knsz03} have proposed that there is no
universal correlation length exponent $\nu$ in the two-dimensional fuse
network.  The numerical evidence presented is based on a disorder 
having a small, finite localization length so that $p$ is not spatially
stationary due
to localization.  However, the analysis implicitly assumes that Eq.\ 
(\ref{xip}) is valid, which requires $p$ to be spatially stationary.
Hence, there is no support for the conclusion reached.

It is the aim of this letter to measure $\nu$ in the three-dimensional
fuse model.  We find the value $\nu=0.86\pm0.06$.  This is close to the
three-dimensional percolation value $\nu=0.88$, hence supporting the notion
that the fuse model is in the universality class of ordinary percolation, 
both in two and three dimensions.  The roughness exponent $\zeta$ was
measured by Batrouni and Hansen \cite{bh98} to be  $\zeta=0.62\pm 0.05$.
Using Eq.\ (\ref{zetanu}) with $\nu=0.86$, we find $\zeta=0.63$.  Hence,
the value for $\nu$ we report here is consistent with the roughness
exponent measured in \cite{bh98} when using Eq.\ (\ref{zetanu}).   
We note, however, that this value for
the roughness exponent is not consistent with the one reported by 
R{\"a}is{\"a}nen et al.\ \cite{rsad98}, who reported a roughness exponent 
close to the minimal energy result, $\zeta=0.41\pm0.02$ \cite{ad96}, claiming 
that they should be identical.

\begin{figure}[t!]
\begin{minipage}{0.46\linewidth}
\centering
\epsfig{file = 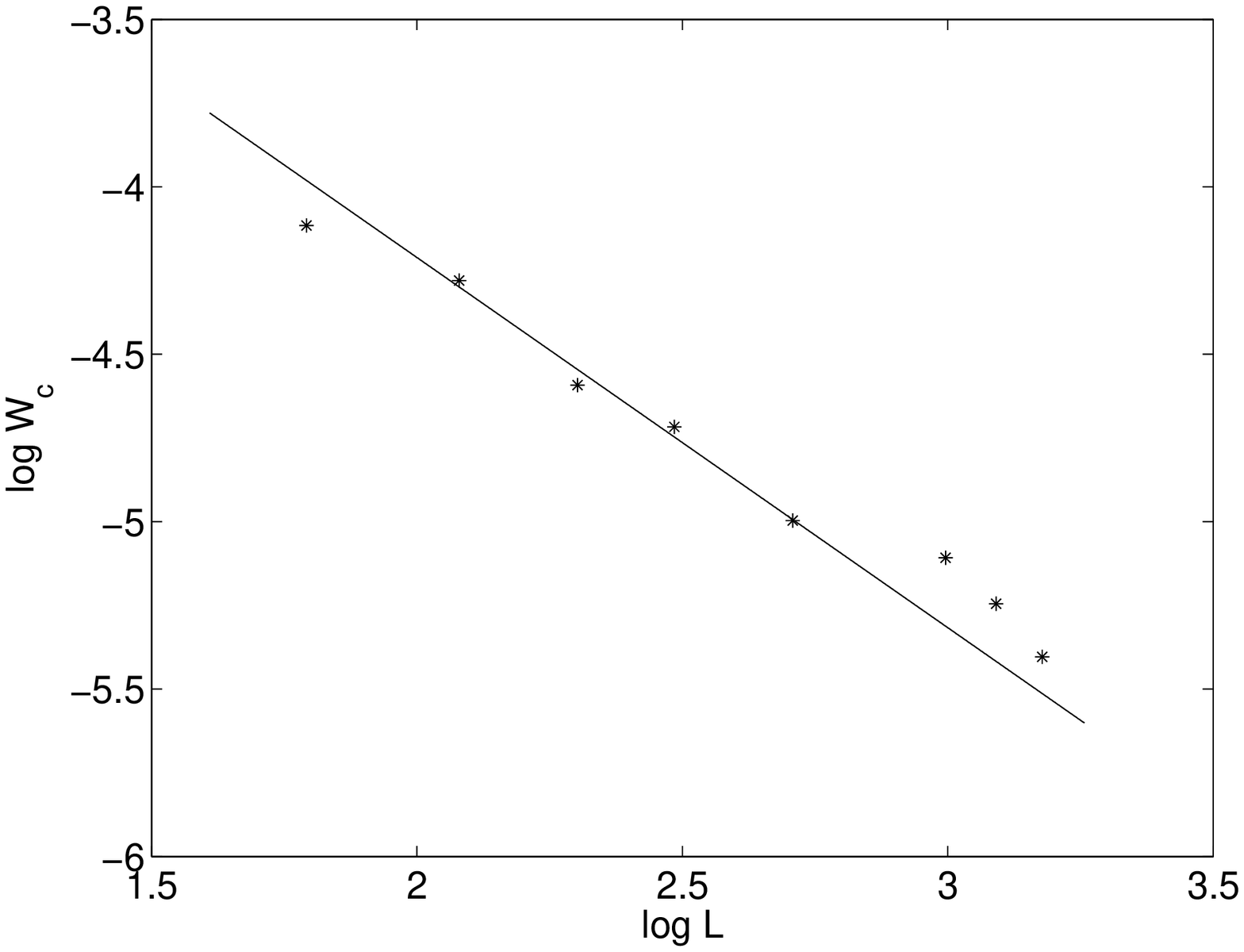, height = 4.5 cm, width = \linewidth, clip=}
\end{minipage}
\hfill
\begin{minipage}{0.46\linewidth}
\centering
\epsfig{file = 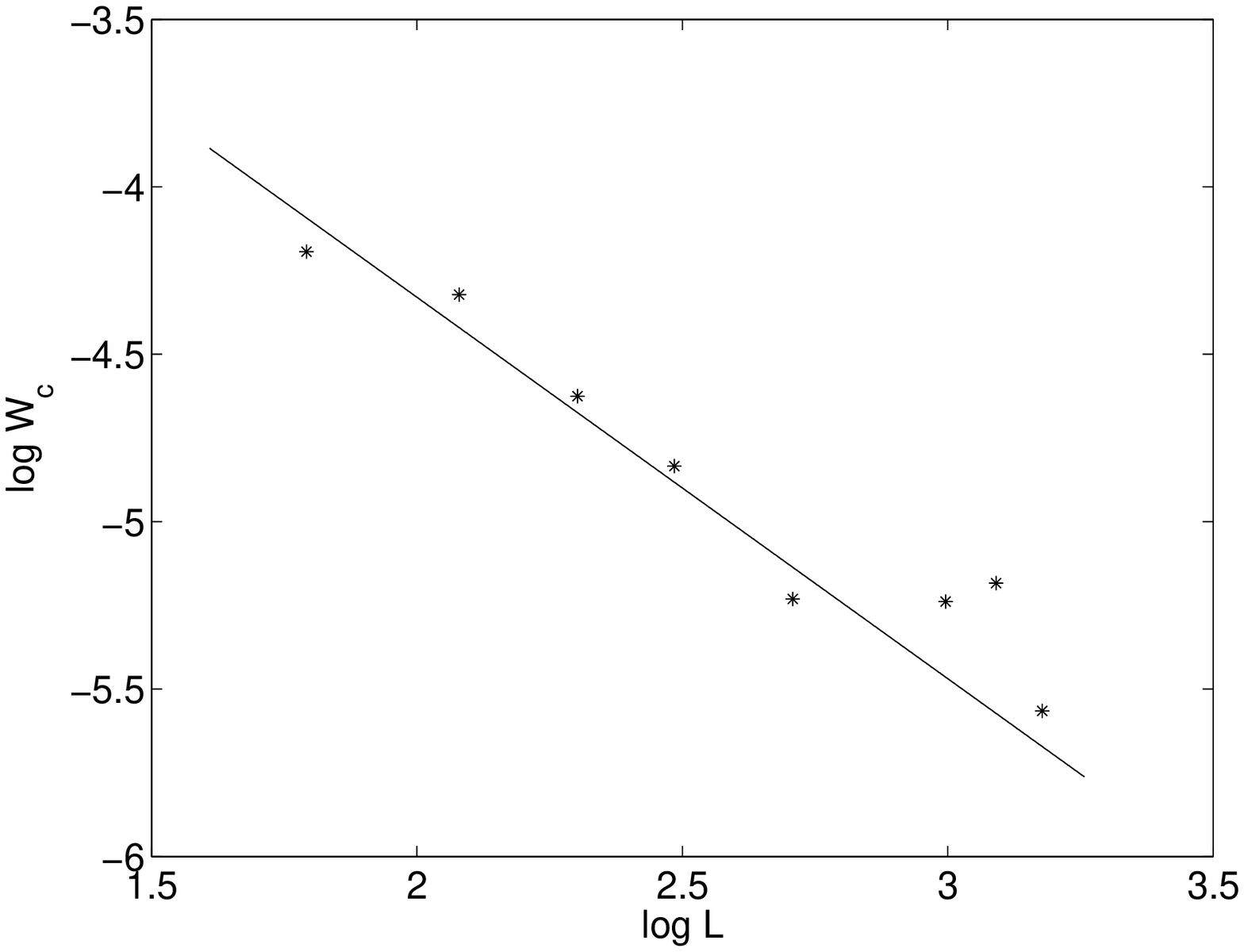, height = 4.5 cm, width = \linewidth, clip=}
\end{minipage}
\hfill
\begin{minipage}{0.46\linewidth}
\centering
\epsfig{file = 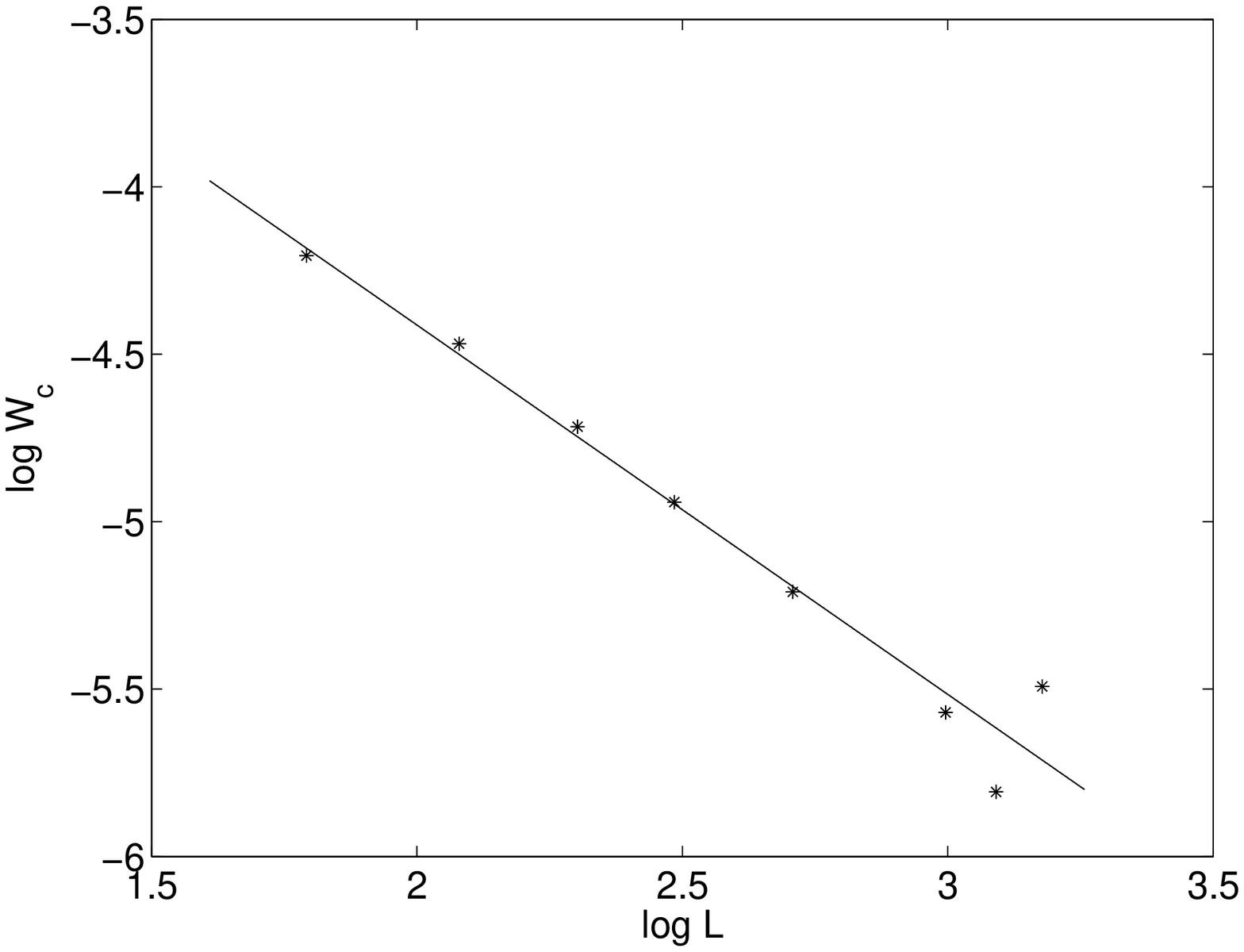, height = 4.5 cm, width = \linewidth, clip=}
\end{minipage}
\hfill
\begin{minipage}{0.46\linewidth}
\centering
\epsfig{file = 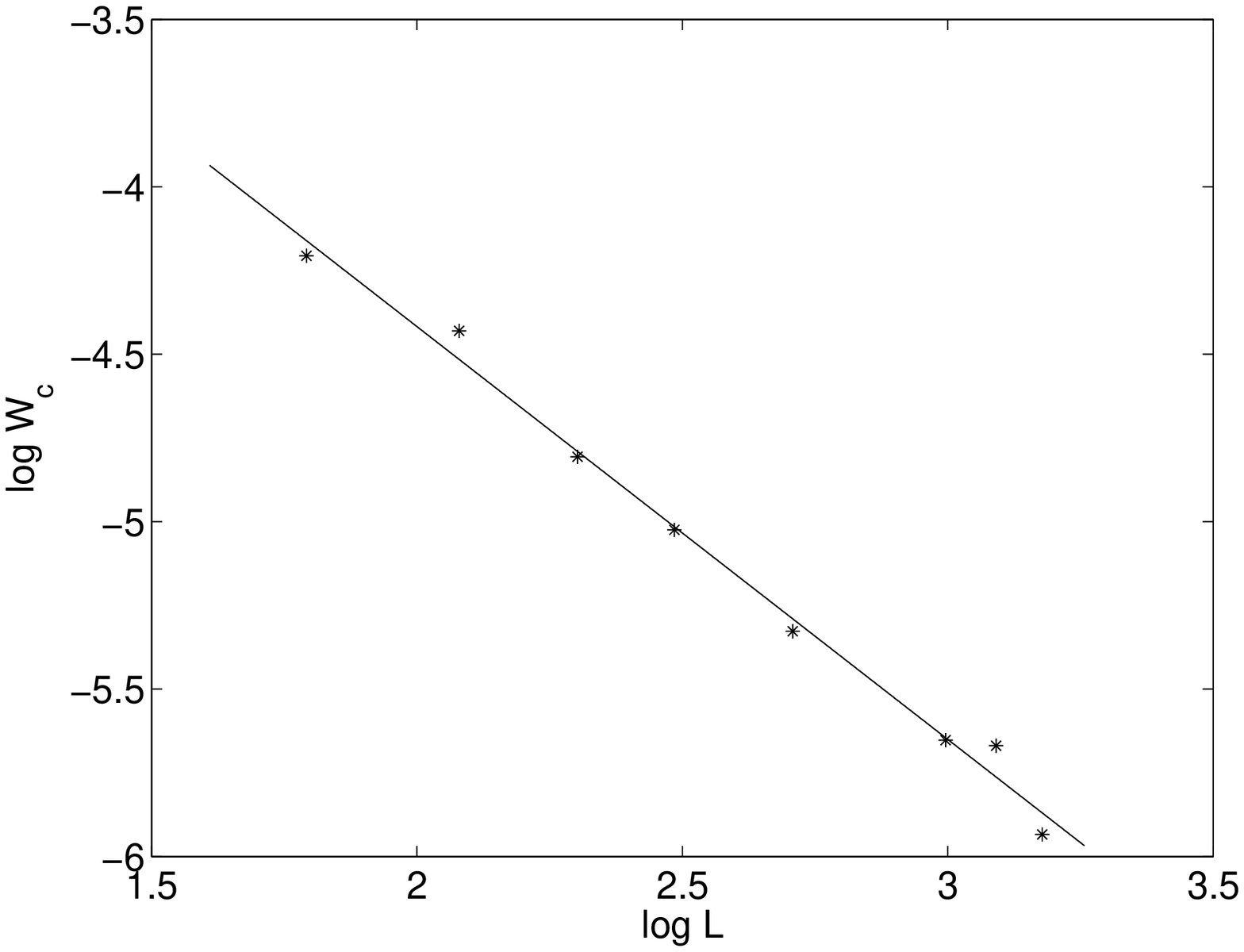, height = 4.5 cm, width = \linewidth, clip=}
\end{minipage}
\caption{Log-log plot of the fluctuations of the density of broken bonds 
$W_c = (\langle p^2 \rangle - \langle p \rangle^2)^{1/2}$ against $L$. 
The disorder is hence $D$ = 10, 12, 15 and 20 respectively.  The slopes
are for $D=10$: 1.11, $D=12$: 1.14, $D=15$: 1.10 and $D=20$: 1.23.  Their mean
is $1/\nu = 1.16\pm 0.06$, giving $\nu=0.86\pm0.06$.}  
\label{fig2}
\end{figure}
The fuse model that we study consists of an oriented simple cubic lattice.
As in Ref.\ \cite{bh98}, we use {\it periodic boundary conditions 
in all directions\/} \cite{nb99} and the average current 
flows in the (1,1,1)-direction. Each bond is an ohmic resistor up to a 
threshold value. When this value is reached, the resistor turns 
irreversibly into an insulator.  The threshold values are drawn from a 
spatially uncorrelated probability density $p(t)$.  A voltage drop equal to 
unity is set up across the lattice along a given plane orthogonal to the 
(1,1,1)-direction.  The currents are then calculated using the Conjugate
Gradient algorithm \cite{bh88}.  After the currents $i$ have been determined,
the bond having the largest ratio $\max(i)/\max(t)$ is determined.  This
bond is then removed and the currents are recalculated.
We do not allow the final crack to cross the plane along which the voltage
drop is imposed. This simplifies the analysis of the final crack breaking the
network apart, while it only imposes weak finite size corrections to fracture
patterns.

The threshold values $t$ constructed by setting $t=r^D$, where $r$ is
drawn from a uniform distribution on the unit interval \cite{bhr94}.  
This corresponds to a probability density $p(t)  \propto t^{-1+\beta}$ on
the interval $0 < t < 1$ with 
$\beta = 1/D$.  The parameter $D>0$ controls the width of the distribution: 
Larger values of $D$ corresponds to stronger disorder.  In order to ensure that
our results are obtained in the strong disorder phase of the fuse model, we
studied $D=$ 10, 12, 15 and 20. Our system sizes varied from $L=6$ to 24 with
5000 samples generated for the smallest sizes to 200 samples for the largest
sizes.

With $D=20$, the smallest threshold values generated are of the order 
$(24^3)^{-20}\approx 10^{-83}$.  The system has, however, still not entered 
purely screened percolation regime.  With this level of disorder, the system
fails when a fraction of about 0.62 of the bonds have failed.  The threshold
values of the bonds that fail near the end of the process are about 
$0.62^{20} \approx 10^{-4}$ --- which is of the order of the currents that
are carried by the bonds in the system.  Hence, there is competition between
threshold values and currents, making the failure process a correlated one
rather than a pure percolation one even in this seemingly extreme case.

\begin{figure}[t!]
\centering
\epsfig{file = 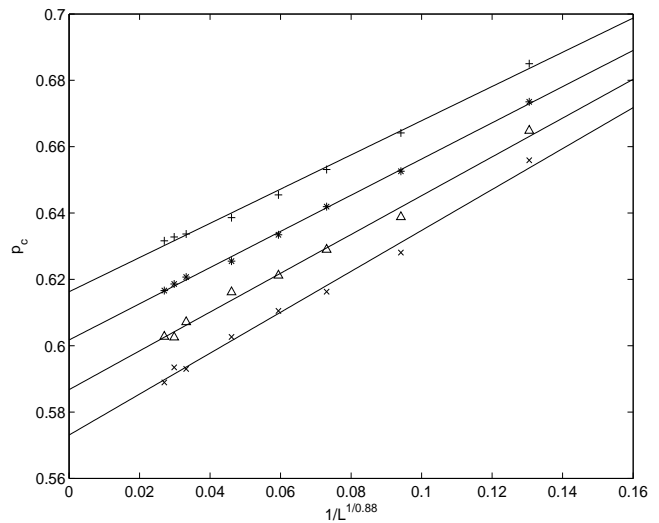, height = 7.0 cm, width = \linewidth, clip=}
\caption{$\langle p \rangle$ plotted against $L^{-1/\nu}$ with $\nu = 0.88$. 
$D = 10 \ (\times)$, $D = 12 \ (\triangle)$, $D = 15 \ (*)$ and $D = 20 \ (+)$.
As $L \rightarrow \infty$ the straight lines extrapolate to the thresholds 
$p_c = 0.573$, $p_c =0.587$, $p_c = 0.602$ and $p_c = 0.616$ respectively.}
\label{fig3}
\end{figure}

Fig.\ \ref{fig1} shows the damage profile in the current direction of the 
random fuse model with $D = 10$.  We denote the (1,1,1)-direction the 
$z$-direction.  We define the damage as the normalized average number of
burned-out fuses in the plane orthogonal to the $z$-direction at $z$.  The
distribution has a weak maximum in the middle.  This indicates a finite but
large localization length $l$.  Such a maximum is smaller or entirely absent 
from the stronger disorders (i.e.\ larger $D$-values) we studied.

Following percolation analysis \cite{sa92}, we define the survival 
probability $\Pi$ indicating the relative number of lattices that has 
survived for a given average damage $p$.  Assuming that the disorder is broad 
enough so that $p$ is independent of $z$ and there is a finite critical value 
of $p=p_c$ at which 50 \% of the lattices survives, we have that
\begin{equation}
\Pi = \Pi[(p - p_c)L^{1/\nu}] \;.
\label{pi}
\end{equation}
This scaling ansatz implies that both the mean value of the density of 
broken bonds $\langle p \rangle$ and the fluctuations 
$(\langle p^2 \rangle - \langle p \rangle^2)^{1/2}$ at breakdown 
scales as $L^{-1/\nu}$ using
\begin{equation}
\langle p \rangle = \int p\biggl(\frac{d\Pi}{dp}\biggr)dp \;,
\label{average}
\end{equation}
and
\begin{equation}
\langle p^2 \rangle - \langle p \rangle^2 = 
\int (p - \langle p \rangle)^2\biggl(\frac{d\Pi}{dp}\biggr)dp \;.
\label{std}
\end{equation}

\begin{figure}[t!]
\centering
\epsfig{file = 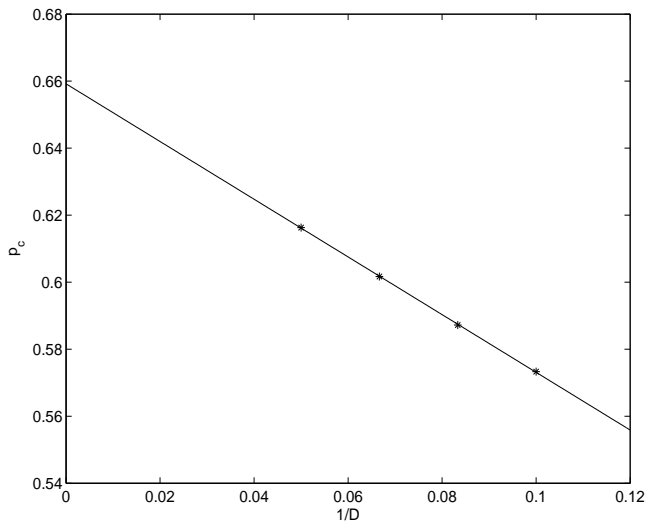, height = 7.0 cm, width = \linewidth, clip=}
\caption{$p_c$ plotted against $1/D$  and extrapolated to 
infinite disorder giving $p_c(\infty) = 0.66$. Extrapoating the straight line,
towards increasing $1/D$-values, we find that $p_c$ becomes negative for
$1/D > 0.75$.}  
\label{fig4}
\end{figure}

In Fig.\ \ref{fig2} the fluctuations of the density of broken bonds, 
$W_c = \sqrt{\langle p^2 \rangle - \langle p \rangle^2}$, have been plotted 
against the system size $L$.  The mean value of the slopes gives 
$\nu = 0.86 \pm 0.06$ which is consistent with percolation value 
$\nu = 0.88$.

Using $\nu = 0.88$ from standard percolation we now turn to the scaling of 
$\langle p \rangle$. From finite-size scaling analysis, we expect the 
functional dependency
\begin{equation}
\langle p \rangle = p_c - \frac{A}{L^{1/\nu}}
\end{equation}
on $L$.  We show this relation for different values of $D$ in Fig.\ 
\ref{fig3}. 

This way of measuring the critical exponent $\nu$ is much less sensitive than 
the one presented in Fig.\ \ref{fig2}.  From standard percolation in a 
simple cubic lattice the threshold for an infinite system is $p_c = 0.752$ 
\cite{sa92}. The extrapolations done in Fig.\ \ref{fig4} show results lying 
below this threshold. However, this is to be expected 
as the percolation process in this limit is {\it screened\/} \cite{rhhg88}.
This result strongly indicates that there is a strong disorder regime for
finite disorders with $p_c$ larger than zero in the three-dimensional fuse 
model.  In fact, extrapolating the straight line in Fig.\ \ref{fig4} towards
larger $1/D$-values will result in $p_c$ reaching zero and becoming negative 
at $D<1.33$.  This is physically impossible and $p_c$ remains zero in this 
range.  This indicates that there is a transition from a percolation-like
regime with $p_c>0$ for $D>1.33$ to a regime with $p_c=0$ for $D<1.33$.
This latter regime has been described as the diffuse localization regime in
\cite{hhr91}.

In summary, we have determined the correlation length exponent in the 
three-dimensional fuse model to be $\nu = 0.86 \pm 0.06$. This is consistent 
with the percolation value of $\nu = 0.88$.  Furthermore, using Eq.\ 
(\ref{zetanu}), this is consistent with the previously measured roughness
exponent $\zeta=0.62\pm0.05$ \cite{bh98}, lending support to the scenario 
proposed
by Hansen and Schmittbuhl \cite{hs03} for understanding the universality of
the roughness exponent in the fuse model and brittle fracture.  Our analysis
was based on studying the fuse model with strong enough disorder for the 
breakdown process to develop in a percolation-like manner with $p$ 
spationally stationary so that the tools developed for studying that 
problem could be used 
in the present one.  We note that in this regime, one will not see the fracture
roughness scaling of Eq.\ (\ref{eq1}):  The fracture will have a fractal
structure.  When, on the other hand, the disorder is weak enough for 
localization to set in, $p$ is no longer spatially stationary, making a 
direct measurement of $\nu$ based on fluctuations in $p$ impossible.  
However, it is in this
regime fracture roughness scaling as in  Eq.\ (\ref{eq1}) is seen as shown in
\cite{bh98}.  

We thank G.\ G.\ Batrouni, H.\ A.\ Knudsen, and J.\ Schmittbuhl
for helpful discussions. B.\ Skaflestad is also greatly acknowledged for giving
help and hints during the implementation of the numerical simulations.


\end{document}